\documentclass[twocolumn,english,aps,prl,reprint,superscriptaddress]{revtex4-1}
\usepackage[LGR,T1]{fontenc}
\usepackage[latin9]{inputenc}
\usepackage{verbatim}
\usepackage{amstext}
\usepackage{amssymb}
\usepackage{graphicx}
\usepackage[version=3]{mhchem}
\usepackage{braket}
\newcommand{\Mn}[1]{\ce{\{Mn#1}\}}

\makeatletter

\ProvideTextCommand{\~}{LGR}[1]{\char126#1}

\@ifundefined{textcolor}{}
{%
 \definecolor{BLACK}{gray}{0}
 \definecolor{WHITE}{gray}{1}
 \definecolor{RED}{rgb}{1,0,0}
 \definecolor{GREEN}{rgb}{0,1,0}
 \definecolor{BLUE}{rgb}{0,0,1}
 \definecolor{CYAN}{cmyk}{1,0,0,0}
 \definecolor{MAGENTA}{cmyk}{0,1,0,0}
 \definecolor{YELLOW}{cmyk}{0,0,1,0}
}

\usepackage[english]{babel}
\def\urlprefix{}
\def\url#1{}
\addto\captionsenglish{%
}

\usepackage[none]{hyphenat}

\newcommand{\angstrom}{\mbox{\normalfont\AA}}

\makeatother

\usepackage{babel}
\begin{document}

\title{Mn84CaltechUF}

\title{Exploring the magnetic properties of the largest single molecule magnets}
\author{Henry F. Schurkus}
\thanks{Henry F. Schurkus and Dianteng Chen contributed equally to this work.}
\affiliation{Division of Chemistry and Chemical Engineering, California Institute of Technology, Pasadena CA 91125}
\author{Dianteng Chen}
\thanks{Henry F. Schurkus and Dianteng Chen contributed equally to this work.}
\affiliation{Department of Physics, University of Florida, Gainesville, FL 32611}
\author{Matthew J. O'Rourke}
\affiliation{Division of Chemistry and Chemical Engineering, California Institute of Technology, Pasadena CA 91125}
\author{Hai-Ping Cheng}
\affiliation{Department of Physics, University of Florida, Gainesville, FL 32611}
\author{Garnet K.-L. Chan}
\affiliation{Division of Chemistry and Chemical Engineering, California Institute of Technology, Pasadena CA 91125}
\date{\today}

\begin{abstract}
  The giant \Mn{70} and \Mn{84} wheels are the largest nuclearity single-molecule magnets synthesized to date
  and understanding their magnetic properties poses a challenge to theory. Starting
  from first principles calculations, we explore the magnetic properties and excitations
  in these wheels using effective spin Hamiltonians. We find that the unusual geometry of the superexchange
  pathways leads to weakly coupled \Mn{7} subunits carrying an effective $S=2$ spin.
  The spectrum exhibits a hierarchy of energy scales and massive degeneracies, with the lowest energy excitations arising
  from Heisenberg-ring-like excitations of the \Mn{7} subunits around the wheel, at energies consistent with
  the observed temperature dependence of the magnetic susceptibility. We further suggest an important role for weak longer-range couplings
  in selecting the precise spin ground-state of the Mn wheels out of the nearly degenerate ground-state band.
  \end{abstract}

\maketitle

Single-molecule magnets (SMM) are fascinating playgrounds in which to probe magnetism at the nanoscale via
targeted chemical design~\cite{christou2000single,sessoli1993high,sessoli1993magnetic}. In this context, 
the giant \Mn{84} wheel was first reported~\cite{tasiopoulos2004giant} in 2004
and more recently, a set of \Mn{70} analogs was synthesized by the same group~\cite{vinslava2016molecules}.
Low temperature magnetic studies suggest ground-state SMM behaviour with
proposed ground-state spins of 6 for \Mn{84} and
5, 7, or 8 for \Mn{70} (depending on the ligand architecture and experimental analysis)~\cite{vinslava2016molecules}, making
these the largest nuclearity SMMs to date.

SMMs of this complexity present a new challenge for theoretical understanding.
The Mn(III) ions in the wheels each carry $S=2$ spin, thus the maximal spin in
\Mn{84} could be as large as 168 if all ions 
were ferromagnetically aligned! However, the Mn ions are bridged by oxygen atoms in a relatively linear configuration, which
gives rise to antiferromagnetic coupling. From this, we would expect a zero or small ground-state spin, but predicting the precise ground-state spin is difficult using elementary arguments.

In a coarse-grained picture, the Mn wheels form a 1D magnetic ring. Coupled 1D quantum spin chains and rings
(that is, chains with periodic boundary conditions) are cornerstones of 1D theoretical quantum magnetism.
However, while phase diagrams of many 1D spin chains are well characterized, including with
respect to non-nearest-neighbor couplings and anisotropy~\cite{golinelli1992haldane,owerre2014haldane}, the focus 
has typically been on the thermodynamic limit where, for example, one can find in
integer (e.g. $S=2$) spin chains, the famous Haldane phase with a finite excitation gap~\cite{schollwock1995haldane}.
Studies that are primarily focused on the behaviour of large, but still finite, magnetic chains and similar structures are much less common,
with the notable exceptions of some studies motivated by SMMs, for example, of
the smaller \Mn{12} wheel~\cite{regnault2002exchange,chaboussant2004exchange,bagai2009drosophila,park2004properties},
or the $\ce{\{X30\}}$ giant Keplerates (X being the magnetic ion)~\cite{muller2001classical,exler2003evaluation,neuscamman2012correlator,ummethum2013large}. The much higher nuclearity of the Mn wheels,
together with the non-trivial arrangement at the atomic level, thus poses a substantial increase in complexity from earlier
theoretical analyses.

In the current work, we will attempt to build a theoretical model of the magnetism in the Mn wheel SMMs. We will first derive
an effective nearest-neighbor magnetic Hamiltonian starting from density functional calculations~\cite{kohn1965self} (DFT) on the \Mn{84} wheel.
Using this magnetic Hamiltonian, we will then use semi-analytic techniques to explore the corresponding low energy states, leading to a model
spectrum for the full wheel. We will see that the particular geometry of the interactions, together with the nearest-neighbor interactions,
yields a strong hierarchy of energy scales and distinct excitation bands with large degeneracies and a very small ground-state gap,
as hinted at in the experimental magnetization data. Because of the large effective degeneracies (including in the ground-state) precise features of the spectrum, such as the
exact ground-state spin, remain sensitive to even weaker long-range couplings. We finish by discussing the limits of the model
and new features that can arise when such long-range couplings are included.

\begin{figure}
  \includegraphics[width=1\columnwidth]{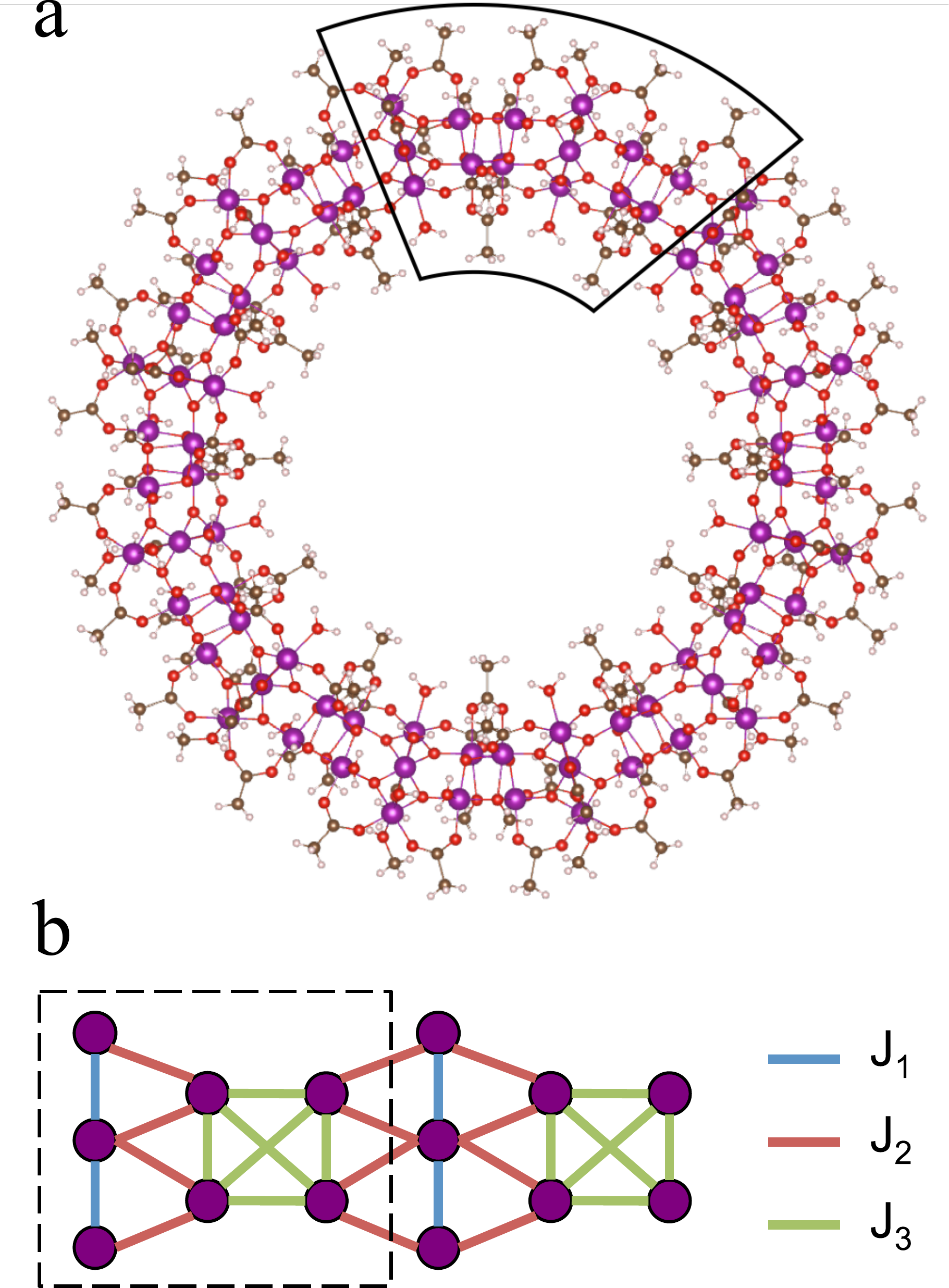}
  \caption{(a) The structure of the
    \Mn{84} wheel. The black frame shows the repeating \Mn{14} unit that
    represents the contents of the asymmetric unit. Colors: Mn purple; O red; C
    brown; H pink
    (b) Schematic of the \Mn{14} and \Mn{7} unit (indicated by dashed box) and dominant super-exchange pathways with exchange
  coupling constants $J_1$, $J_2$, $J_3$.}
  \label{fig:Mn84structure}
\end{figure}

\noindent \paragraph{Structure}
The \Mn{70} and \Mn{84} wheels consist of repeating units of 14 oxo-bridged
Mn(III) ions, illustrated in Fig.~\ref{fig:Mn84structure}(a).
The 14 Mn cell consist of two 3 Mn subunits (``lines'') and two 4 Mn subunits
(``tetrahedra''). We can group
these into two 7 Mn subunits, and in a coarse grained magnetic model (where the
oxygens do not appear, Fig.~\ref{fig:Mn84structure}(b))
we can use these as building blocks of the wheel. There are in fact multiple
ways to choose \Mn{7} cells; for example, we could group the 3 Mn line
with half of each of the tetrahedrons on either side of it as an alternative
\Mn{7} cell; however the structure of the couplings
means that such choices do not greatly affect our analysis below.

\noindent \paragraph{First-principles calculations} Because the \Mn{70} and \Mn{84} wheels contain very similar
structural \Mn{14} and \Mn{7} subunits, we will focus on deriving a model Hamiltonian for the \Mn{84} wheel only.
We carried out $\Gamma$ point first principles Kohn-Sham DFT ~\cite{kohn1965self} calculations on the \Mn{84} wheel
using the spin-polarized Perdew-Burke-Ernzerhof (PBE) exchange correlation functional~\cite{perdew1996generalized} and projector-augmented-wave (PAW) pseudopotentials~\cite{blochl1994projector,kresse1999ultrasoft} in conjunction with the plane-wave basis as implemented in the Vienna Ab-initio Simulation Package (VASP)~\cite{kresse1996efficiency,kresse1996efficient}. We used a
plane-wave cutoff energy of 500 eV with an energy threshold for self-consistency
of $10^{-6}$ eV. We first relaxed the
structure for \Mn{84} with a force threshold per atom of less than 0.05 eV/\angstrom, and used the optimized structure to calculate
the total energies of multiple broken-symmetry collinear spin configurations.

\begin{figure}
  \includegraphics[width=1\columnwidth]{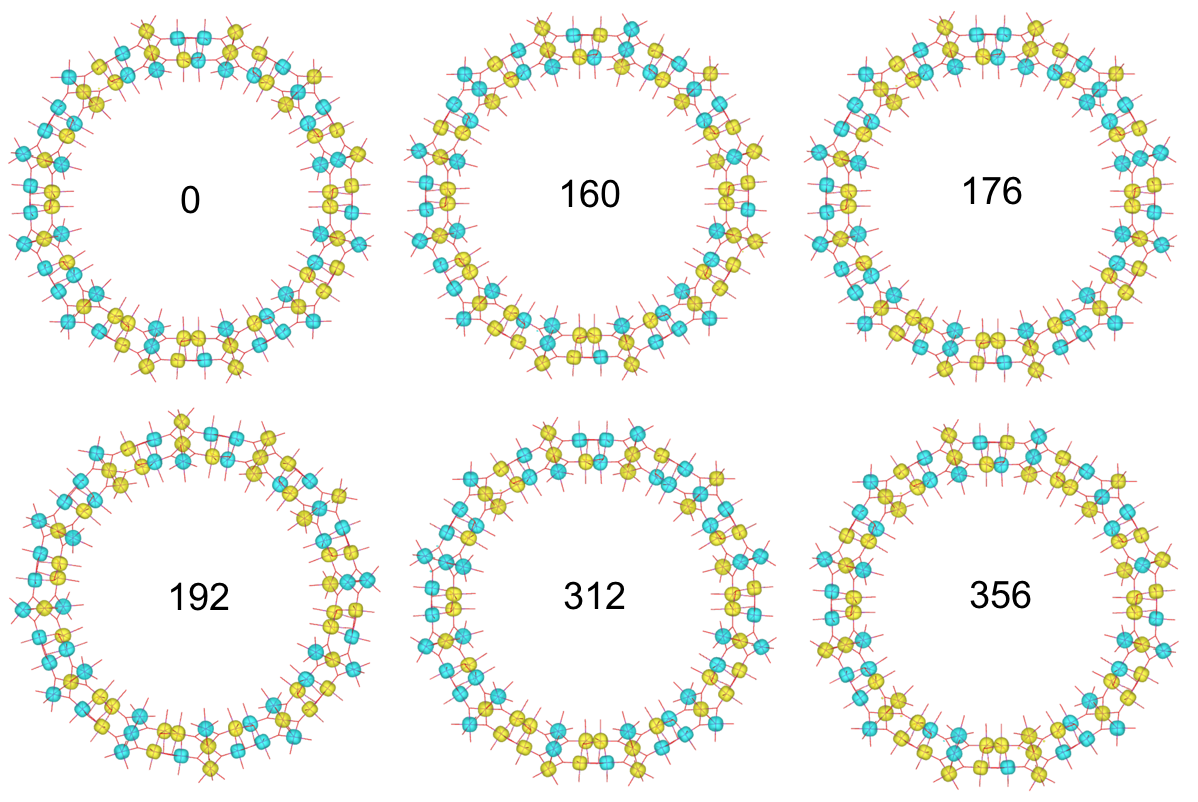}
  \caption{Spin configurations of \Mn{84} used to determine the coupling
    constants $J_1$, $J_2$, and $J_3$. Yellow represents spin up Mn sites ($S^z = +2$),
    cyan represents spin down Mn sites ($S^z = -2$). In the center of each
    spin configuration its total energy in meV is given relative to the lowest
    configuration. The net spins are $S^z_\textrm{tot} = 4, 8, 12$ for
    the top row (from the left to the right), and $S^z_\textrm{tot} = 4, 8, 20$
    for the bottom row, respectively.
  }
  \label{fig:Mn84spinconfigs}
\end{figure}

\noindent \paragraph{Model Hamiltonians} Because the DFT wavefunctions describe collinear spin configurations,
the energies naturally parametrize interactions of the Ising type.
The spin configurations used for the parametrization are shown in Fig.~\ref{fig:Mn84spinconfigs}.
To fit the DFT energies, we used an Ising model with nearest-neighbor interactions
\begin{equation}
    H=\sum_{i<j}J_{ij}{S^z_i}{S^z_j}
\label{eq:H}
\end{equation}
where 
$J_{ij}$ is the exchange coupling. 
Considering only nearest-neighbor interactions, there are three different
exchange interactions in \Mn{84}, shown as $J_1$ $J_2$ and $J_3$ in Fig.~\ref{fig:Mn84structure}.
We obtained coupling constants of $J_{1}$ = 14.2(1.0) meV, $J_{2}$ = 2.3(1.0) meV, $J_{3}$ = 1.0(0.2) meV, where the bracketed numbers
are an estimate of uncertainty. 
We also tried to derive next nearest-neighbor (NNN) interactions, but found it difficult to obtain consistent estimates; calculations 
on a model of the \Mn{14} subunit yielded ferromagnetic NNN couplings, while calculations on \Mn{84} gave antiferromagnetic NNN couplings.
We return to the potential role of long-range couplings in the low-energy states of the Mn wheel in the discussion below.

Although DFT provides information on the Ising part of the interactions, we also consider the case of
isotropic exchange couplings described by the Heisenberg model
\begin{equation}
  H=\sum_{i<j}J_{ij}{\textbf{S}_i}\cdot {\textbf{S}_j}
\end{equation}
We use the same nearest-neighbor $J_{ij}$ as derived for the Ising model (the Heisenberg Hamiltonian yields the same energies as the Ising Hamiltonian
for these couplings given the same broken symmetry spin configurations). The Ising model can be
considered to be the limit of the Heisenberg Hamiltonian either in the case of large anisotropic spin-spin couplings,
or large magnetic ion anisotropy, described by the term $-D\sum_i (S^z_i)^2$.

\begin{figure}
  \includegraphics[width=\columnwidth]{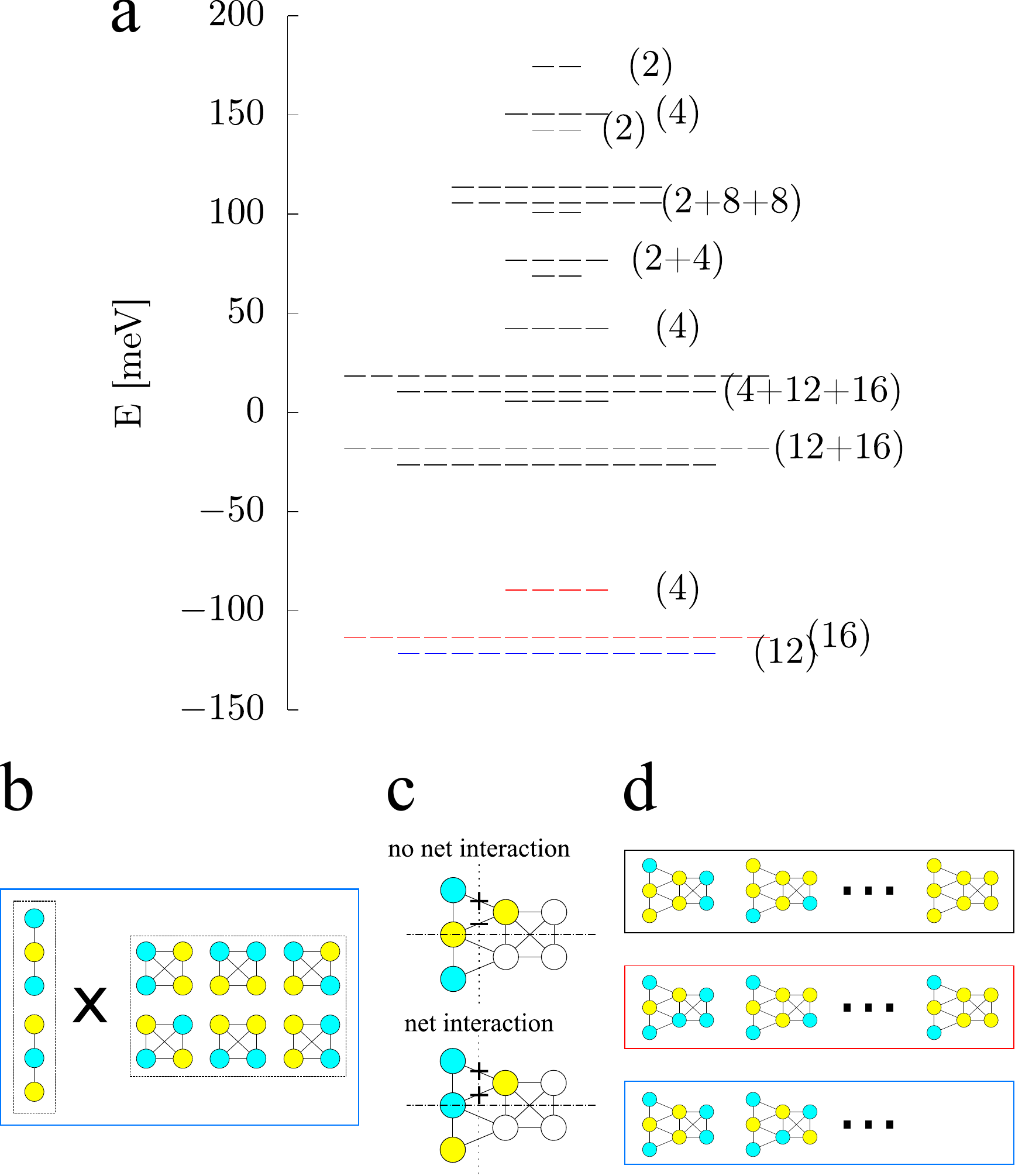}
  \caption{a) Ground- and excited-states for the Ising model representing the
\Mn{7} subunit. Blue: 12 degenerate ground-state levels. Red: Excited states
including only tetrahedron excitations. Black: Excited states including also
line excitations. b) Decomposition of the ground-states into line and
tetrahedron configurations. c) If the line is in its symmetric ground-state,
neighboring tetrahedron sites experience an equal amount of positive and
negative interaction with the neighboring two sites of the line and thus have no
net interaction. If the line is excited an effective interaction ensues.
The lowest such line excitations are asymmetric. d)
Top-to-bottom: Representative spin configurations of the line-excitation band (black box), 
tetrahedron-only excitation band (red box), and ground-state (blue box).}
  \label{fig:IsingModel}
\end{figure}

\begin{figure}[b]
  \includegraphics[width=1\columnwidth]{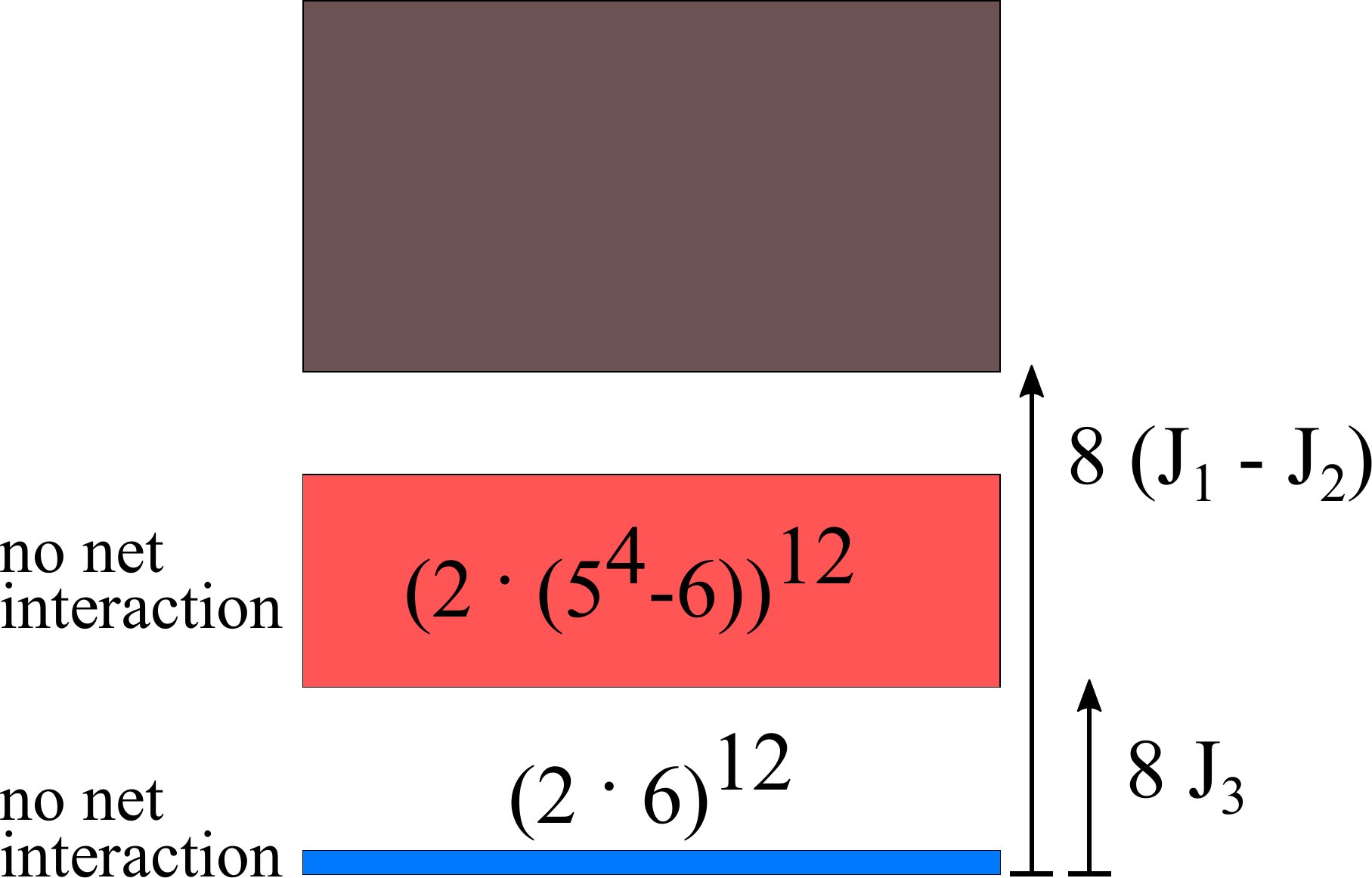}
  \caption{The spectrum of the Ising model corresponding to the \Mn{84} molecule
  decomposes into three bands: the highly degenerate ground-state (blue),
excitations including at least one excited tetrahedron in the ring made up of a
total of 12 tetrahedra and 12 lines, and excitations further including at least
one line excitation.}
\label{fig:Ising84}
\end{figure}

\noindent \paragraph{Analysis: Ising model}
To orient ourselves, we first deduce the spectrum of the Mn wheels
within the simple Ising model.
We start with the \Mn{7} ``unit'' of the ring consisting of the line and tetrahedron. The Ising Hamiltonian then supports 12 degenerate ground-states
shown in blue in Fig.~\ref{fig:IsingModel}(a). We can trace the origin of
the 12 degenerate ground-states to the ground-states of the line and the
tetrahedron (Fig.~\ref{fig:IsingModel}(b)).
The 3 Mn in the line configuration display the largest antiferromagnetic (AFM) exchange coupling, thus the ground-state of the line is an $S^z=2$ configuration, which is doubly degenerate
from the 2 spin orientations.  For the tetrahedron, the equal exchange interactions along each of the
edges leads to magnetic frustration, giving 6 degenerate $S^z=0$ ground-states (corresponding to the 6 different ways
of arranging 2 up and 2 down orientated spins on the tetrahedron). Finally,
because of the exact cancellation of
interactions of the line center site and either end with their tetrahedral
neighbor in the symmetrical ground state (Fig.~\ref{fig:IsingModel}(c)), the interaction energy between the 2 line and 6 tetrahedron ground-states identically vanishes, leading to 12 degenerate ground-states (Fig.~\ref{fig:IsingModel}(b)). 

The low-energy excited-states of the \Mn{7} unit in Fig.~\ref{fig:IsingModel}(a) can similarly be thought of in terms of the excited states of
the line and tetrahedron, as seen in Fig.~\ref{fig:IsingModel}(d). Because of the very different energy scales of the 3 different exchange
interactions $J_1$, $J_2$, $J_3$, the energy spectrum roughly decomposes into 3
sets, the ground-state set (blue levels), tetrahedron excitations starting at $8J_3$ above the
ground-state (red levels) and, line excitations starting at $8(J_1-J_2)$ above the ground
state (black levels). As discussed above, the line and tetrahedron do not interact in the ground-state, and similarly do not interact
in excited states composed of the line ground-state and any of the $2\cdot4$
tetrahedron single or $2$ double spin-flip excitations (thus yielding 16 fold and 4 fold
degeneracy, respectively).
However at energies higher than $8(J_1-J_2)$, one starts to access asymmetric
configurations of the 3 Mn line, and these then have an overall non-zero
coupling to the tetrahedron.

Starting from the \Mn{7} unit, we can also understand the ground-state and excitations of the Mn wheel within an Ising model,
as illustrated in Fig.~\ref{fig:Ising84}.
Since the lowest energy states of the line and tetrahedron do not interact, each 7 Mn unit contributes an effective $S=2$ Ising spin
with 12 fold ground-state degeneracy, and these effective spins do not interact around the ring. This means that the ground-state
of the \Mn{84} wheel acquires a $12^{12}$ fold degeneracy, with total Ising
spins ranging from $S^z_\textrm{tot}=-24$ to $S^z_\textrm{tot}=24$.
A similar picture emerges for the excited states, which incorporate also the excited states of the \Mn{7} units. The lowest
bands of states correspond to sets of \Mn{7} ground-states and at least one \Mn{7} excited state: these do not interact, again due to the lattice symmetry. At higher energies
the possibility for aggregate non-zero interactions between the lines and
tetrahedra is generalized to interactions between the
neighboring \Mn{7} units on the ring.

\noindent \paragraph{Analysis: Heisenberg model} We now turn to the Heisenberg model with the same couplings. The difference
between the quantum Heisenberg model and the classical Ising model is the inclusion of $XX$ and $YY$ couplings which
leads to states where the spins are not maximally aligned
along the spin axis and which are superpositions of classical spin states.

\begin{figure}
  \includegraphics[width=.6\columnwidth]{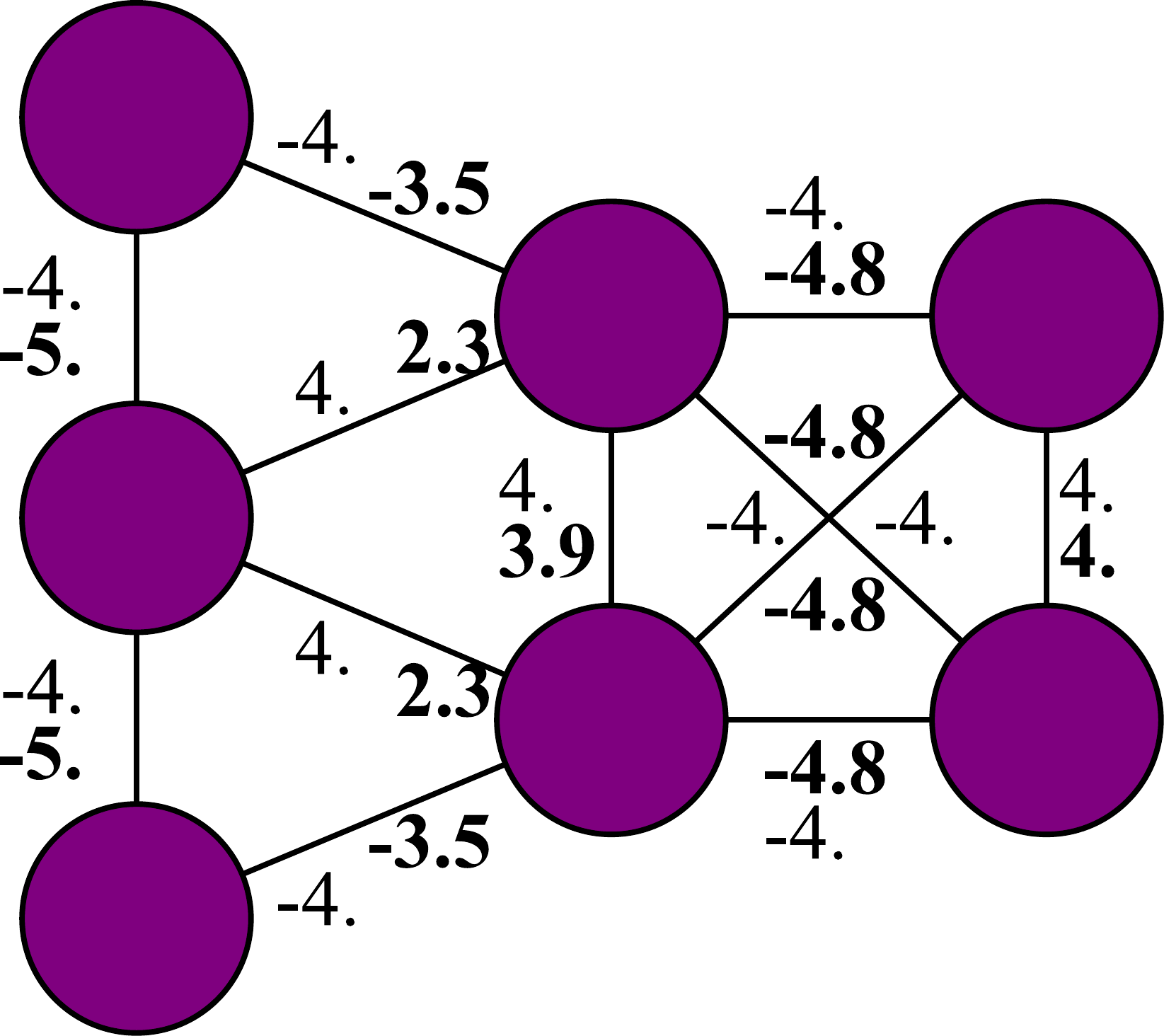}
  \caption{Correlation functions, $\braket{\textbf{S}_i\cdot\textbf{S}_j}$,
  between neighboring spin sites in the ground-state of the Heisenberg model (bold
font) and Ising model (regular font) corresponding to the \Mn{7} subunit. In the
Ising model each spin is $S^z = \pm 2$ and thus the correlation functions equal
$\braket{S^z_i S^z_j}$.}
\label{fig:Correlators7}
\end{figure}

We start with the Heisenberg model for the \Mn{7} ground-state. The ground-state quantum correlation functions
$\braket{\textbf{S}_i \cdot \textbf{S}_j}$ are plotted in Fig.~\ref{fig:Correlators7}
and compared to the Ising correlation functions $\braket{S^z_i S^z_j}$. The similarity in the correlation functions
shows that the quantum ground-state is largely similar in character to that of the classical Ising ground-state.

\begin{figure*}
\includegraphics[width=1\textwidth]{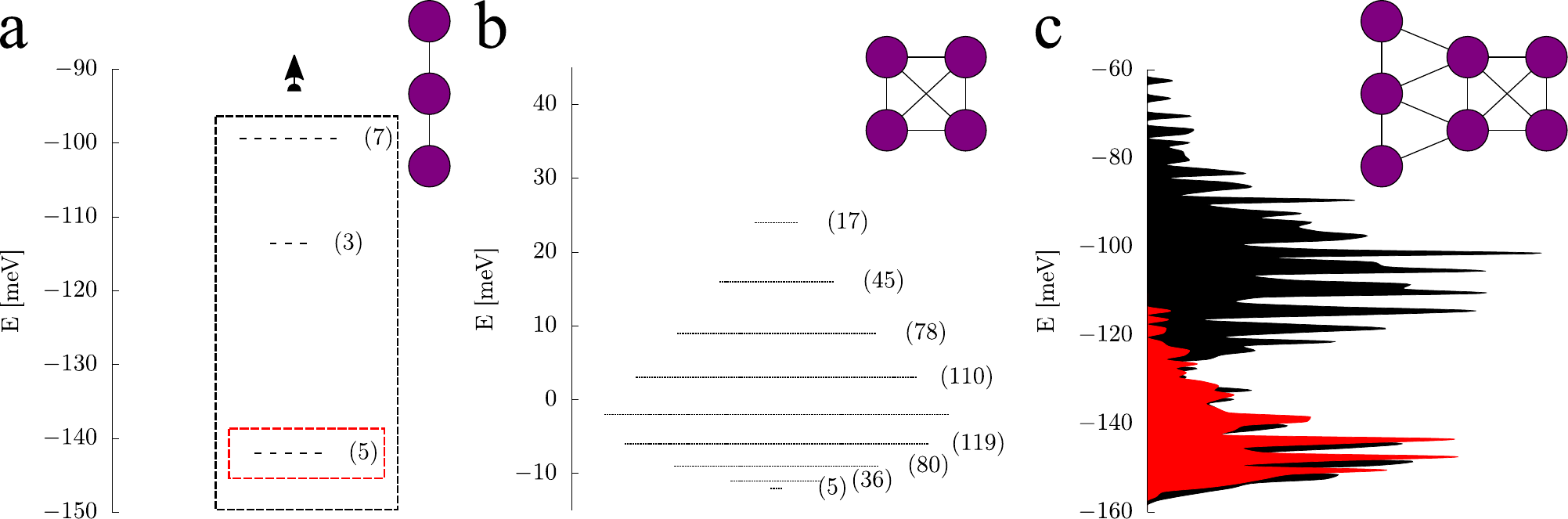}
\caption{Spectrum of the Heisenberg model corresponding to the line
substructure (a) and tetrahedron substructure (b).
Describing the \Mn{7} subunit in the basis of all tetrahedron states combined
with the lowest 15 line states (black box in a) gives a high
quality description of the density of states (black curve in c). Due to the
large relative energy scale of the line interactions, already restricting the
basis in the line substructure to the 5 degenerate ground-states (red box in a),
yields a density of the low energy states (red curve in c) that closely follows
the more accurate one.}
\label{fig:TruncatedSpectrum}
\end{figure*}

As a result of the quantum fluctuations, the energies of the line and tetrahedron are no longer additive, i.e. in the ground-state
the line and the tetrahedron interact with each other, because both the
ground-state and the tetrahedron can now in principle include in their
quantum superpositions excited line configurations which are not the symmetrical
Ising ground-state configurations depicted in Fig.~\ref{fig:IsingModel}.
We compute this interaction energy by exact diagonalization to be -3.44 meV.

Because $J_1$ (the coupling within the line) is much stronger than outside of it, for low energy properties we can
simplify the Hilbert space of the \Mn{7} unit by restricting the line to its lowest states, such as
the (degenerate) $S=2$ ground-state. Truncating the line to its ground-state, the interaction energy
in the ground-state between the line and tetrahedron is then computed to be -2.46 meV;
truncating the line to lowest 2 and 3 sets of spin-states in the line gives -3.00 meV and 
-3.43 meV respectively, very close to the result obtained by exact diagonalization.
The quantum spectrum of the \Mn{7} unit truncating the line to its lowest and three lowest
spin manifolds is shown in Fig.~\ref{fig:TruncatedSpectrum}. This  confirms that the lowest energy part of the spectrum
is qualitatively correct even if we truncate the line only to its ground-state space.
Comparing the spectrum of the Heisenberg model to that of the Ising model in Fig.~\ref{fig:IsingModel}(a) we see echoes of the degeneracies;
although the exact degeneracies are lifted due to the quantum interactions between the line and tetrahedron,
we see a band of spin-states starting at roughly $\mathcal{O}(J_3)$ above the
ground-state, and a second band of states roughly $\mathcal{O}(J_1-J_2)$ above
the first band.

\begin{figure}
  \includegraphics[width=1\columnwidth]{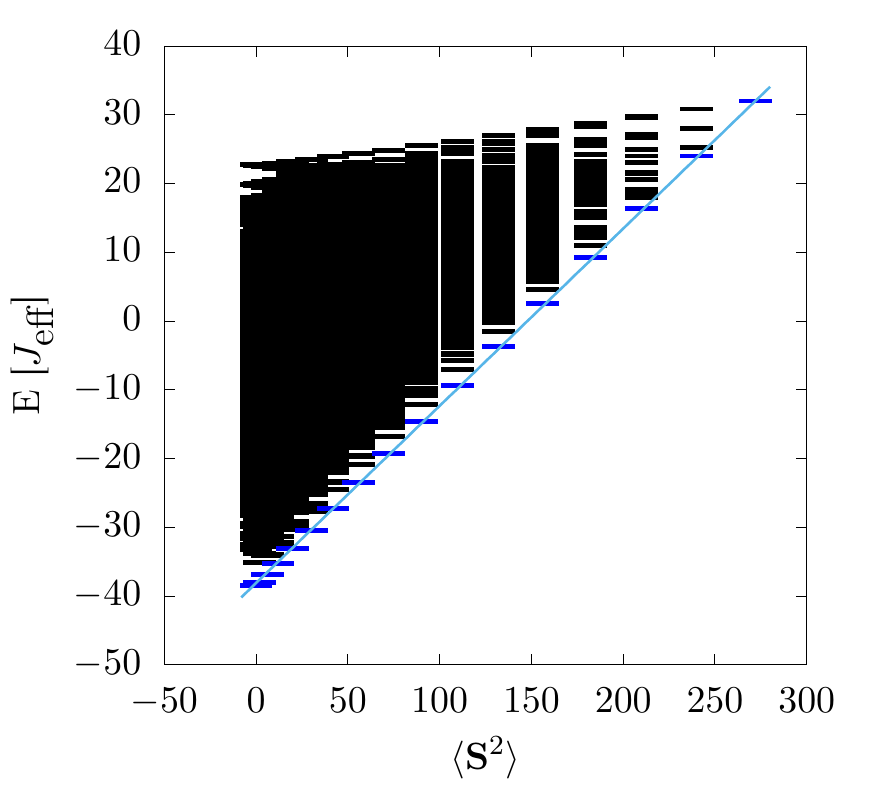}
  \caption{Spectrum of $S=2$ Heisenberg ring model with $L=8$ sites (\Mn{56}). The rotational band model predicts
    $\textrm{E}(\braket{\textbf{S}^2}) \approx
  J_\textrm{eff} \frac{\kappa}{L}\braket{\textbf{S}^2} + E_a$, where $E_a$ is
  determined such that the rotational band model and the spectrum agree exactly
  for the highest
$\braket{\textbf{S}^2}$ state. Fitting yields $\kappa = 2.062$.}
  \label{fig:S2N8}
\end{figure}

Just as quantum fluctuations introduce interactions between the line and the
tetrahedron
in the ground-state of the \Mn{7} unit, quantum fluctuations similarly introduce interactions between the \Mn{7} units, lifting the exact
degeneracies seen in the classical spectrum of the wheel. We can then expect the degenerate blue states in Fig.~\ref{fig:Ising84} to
be replaced by a band of states.
If we view each \Mn{7} unit as an effective $S=2$ unit, we can write
an effective Heisenberg model for interactions between the \Mn{7} unit,
$H_\text{eff} = \sum_{ij} J_\text{eff} \textbf{S}_i \cdot \textbf{S}_j$, where
$\textbf{S}_i, \textbf{S}_j$ are
now operators measuring the spin of each \Mn{7} unit. To estimate the effective coupling we have diagonalized the 
10 Mn line-tetrahedron-line unit. Since the Hilbert space is somewhat large for
exact diagonalization ($5^{10}$),
we again simplify the problem by building a Hilbert space consisting only of the $S=2$ subspace of each Mn line and the lowest $121$ states (lowest
3 spin manifolds) of the tetrahedron. The ground-state is  $S=0$ and the first excited state is  $S=1$ 
(consistent with the lowest excited states of two coupled $S=2$ effective spins)
and this gives an estimate of an antiferromagnetic $J_\textrm{eff}$ of 0.29 meV.
From this effective antiferromagnetic Heisenberg model, we then obtain that the
ground-state of the ring is overall a singlet state for both \Mn{70} and
\Mn{84}. 
The degenerate ground-states in the classical Ising model are thus replaced by the spin-wave like spectrum of an $L$ site Heisenberg ring.
This is shown in Fig.~\ref{fig:S2N8} for $L=8$ \Mn{7} units (\Mn{56}), where the lowest spin-excitation at each
$S$ value is roughly $J_\textrm{eff} \kappa \braket{\textbf{S}^2}/L$ following the rotational-band
model~\cite{schnack}. The lowest excitation in the \Mn{84} wheel is thus one of the
excitations of the effective $L=12$ Heisenberg ring (exciting from $S=0$ to $S=1$) which appears at about
0.05 meV, or about 0.6 K.

\noindent \paragraph{Discussion} Our above analysis using the nearest-neighbour Ising and Heisenberg models
brings to light some general features of the Mn wheel spectrum. In particular, in the Ising model,
we see that the symmetry of the lattice and interactions between the lines and the tetrahedra
makes them very weakly coupled. Together with the spread in the nearest-neighbour exchange couplings $J_1$, $J_2$, $J_3$,
this gives rise to a hierarchy of energies separating excitations, each of which displays massive exact degeneracies.
For the low energy parts of the spectrum, we can consider the lines to be effective carriers of $S=2$ spins.

The primary effect of quantum fluctuations in the Heisenberg model is to spread
each of the exact degeneracies into a band of states. The effective coupling between the lines together with spin-wave
like excitations around the ring produces a new lower energy scale on the scale of $\sim 1$K, smaller than
any of the nearest-neighbour exchange couplings. The experimental magnetic susceptibility shows a steep change in the range of 2K-10K~\cite{vinslava2016molecules}. This has previously been interpreted as arising from the thermal depopulation of excited state energy levels, and
our model indicates that these energy levels are those associated with the Heisenberg spin-wave excitations of effective $S=2$ subunits around the wheel.

Since many of our deductions arise from symmetry, they are quite insensitive
to the values of $J_1, J_2, J_3$. For example, the ground-state spins and degeneracy in the Ising model are independent
of the magnitude of  $J_1$, $J_2$, $J_3$ and depend only on their signs, and similarly $J_\text{eff}$ in the Heisenberg model
always remains small as fluctuations are needed to induce an overall interaction. Nonetheless, the exact degeneracies in the Ising model
and small $J_\text{eff}$ Heisenberg coupling means that even smaller terms in the Hamiltonian are likely to affect the precise
ordering of states in each of the excitation bands. This includes the ground-state band, where such terms will determine
the exact ground-state spin, which is zero in our model, but observed to be clearly non-zero in experiment.

\begin{figure}
  \includegraphics[width=1\columnwidth]{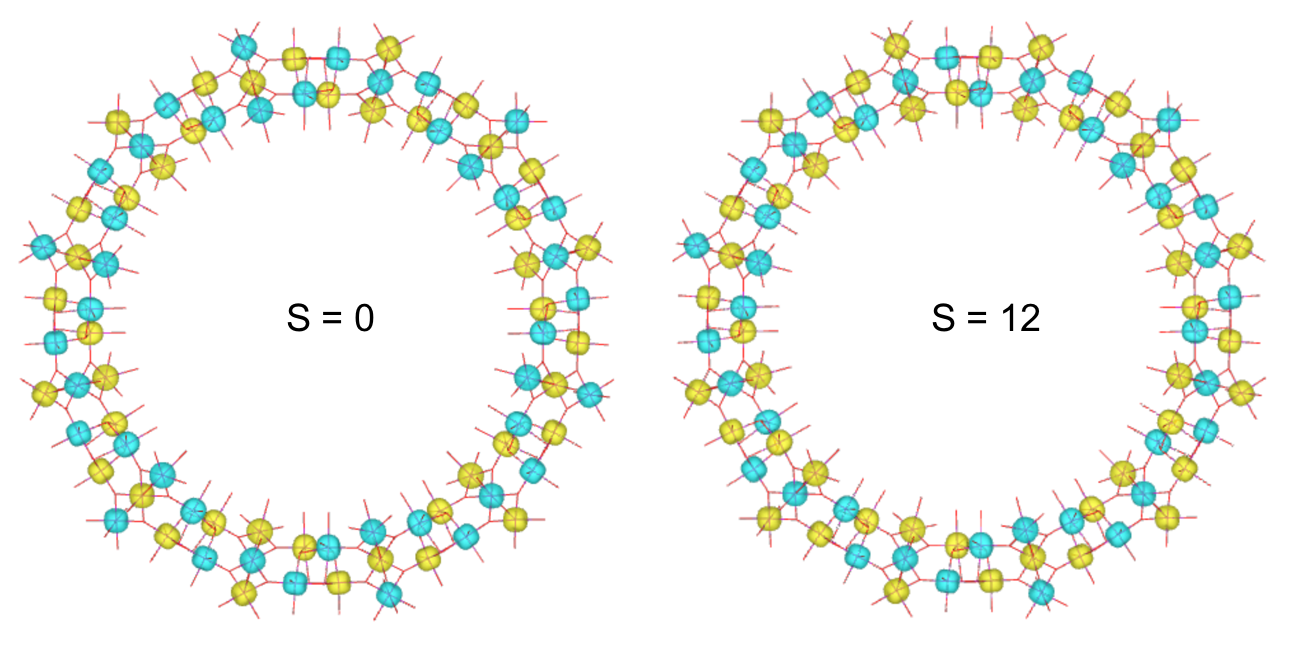}
  \caption{The spin configurations of $S = 0$ and $S = 12$ constructed by up and down $S=2$ \Mn{7} units with the same local spin configurations. The $S = 12$ spin configuration has a total energy 60 meV lower than the $S = 0$ spin configuration.}
  \label{fig:Spin0_and_spin12}
\end{figure}

Because the DFT configurations we used in Fig.~\ref{fig:Mn84spinconfigs}
were chosen to probe local spin flips and we derived a model based on nearest-neighbour couplings, the most important candidate for remaining physics
is the longer range (such as non-nearest-neighbour (NNN)) exchange couplings. Even so, the effects
of non-nearest-neighbour couplings are subtle, as a single  $J_4$ NNN term, given the geometry of the lattice,
will still yield an overall zero ground-state spin in the Ising model. Nonetheless, to explore this, we computed the 
$S^z_\textrm{tot}=0$ and $S^z_\textrm{tot}=12$ DFT ground-states,  keeping identical local spin configurations in the \Mn{7} subunits as shown in
Fig.~\ref{fig:Spin0_and_spin12}, to probe the effects of longer-range couplings. In fact, we
find the $S^z_\textrm{tot}=12$ ground-state to be 60 meV lower than the
$S^z_\textrm{tot}=0$ state, demonstrating that there are subtle longer-range effects. However, we also cannot conclude
that the effective coupling between \Mn{7} units is purely ferromagnetic, as that would lead to an $S=24$ ground-state
for \Mn{84}. Instead, frustration between the short-range effective AFM coupling that we have considered in this work
and some weaker longer-range effective FM couplings will reorder the lowest states in the Heisenberg spectrum in Fig.~\ref{fig:S2N8}. 
For example, for an Ising model with nearest-neighbor interactions
and a $k$th-neighbor interaction $J_k \sum_i S^z_i S^z_{i+k}$, will have in the limit of large $J_k$, every $k$th spin either
ferromagnetically or antiferromagnetically aligned. Of course, more complicated mechanisms involving frustrating interactions at different ranges (rather than a single one at range $k$) will be more natural. The fact that the ground-state spin in the Mn wheels is approximately equal
to the number of \Mn{14} subunits suggests that the longer-range interactions can perhaps be coarse grained into
an effective ferromagnetic Heisenberg model involving the \Mn{14} units. However, obtaining a microscopic Hamiltonian
that correctly incorporates all these small energy scales remains challenging and remains a topic for further work.

\paragraph{Acknowledgements}
This work was supported by the US Department of Energy, Office of Science, via the M$^{2}$QM EFRC under award no. DE-SC0019330.
We thank Ashlyn Hale for help with the \Mn{84} crystal structures.
HFS acknowledges funding from the European Union's Framework Programme for Research and Innovation Horizon 2020 (2014-2020) under
the Marie Sk\l{}odowska-Curie Grant Agreement No. 754388 and from LMUexcellent as part of LMU Munich's funding as a University of Excellence within the framework of the German Excellence Strategy.  

\bibliography{References}

\begin{thebibliography}{23}%
\makeatletter
\providecommand \@ifxundefined [1]{%
 \@ifx{#1\undefined}
}%
\providecommand \@ifnum [1]{%
 \ifnum #1\expandafter \@firstoftwo
 \else \expandafter \@secondoftwo
 \fi
}%
\providecommand \@ifx [1]{%
 \ifx #1\expandafter \@firstoftwo
 \else \expandafter \@secondoftwo
 \fi
}%
\providecommand \natexlab [1]{#1}%
\providecommand \enquote  [1]{``#1''}%
\providecommand \bibnamefont  [1]{#1}%
\providecommand \bibfnamefont [1]{#1}%
\providecommand \citenamefont [1]{#1}%
\providecommand \href@noop [0]{\@secondoftwo}%
\providecommand \href [0]{\begingroup \@sanitize@url \@href}%
\providecommand \@href[1]{\@@startlink{#1}\@@href}%
\providecommand \@@href[1]{\endgroup#1\@@endlink}%
\providecommand \@sanitize@url [0]{\catcode `\\12\catcode `\$12\catcode
  `\&12\catcode `\#12\catcode `\^12\catcode `\_12\catcode `\%12\relax}%
\providecommand \@@startlink[1]{}%
\providecommand \@@endlink[0]{}%
\providecommand \url  [0]{\begingroup\@sanitize@url \@url }%
\providecommand \@url [1]{\endgroup\@href {#1}{\urlprefix }}%
\providecommand \urlprefix  [0]{URL }%
\providecommand \Eprint [0]{\href }%
\providecommand \doibase [0]{http://dx.doi.org/}%
\providecommand \selectlanguage [0]{\@gobble}%
\providecommand \bibinfo  [0]{\@secondoftwo}%
\providecommand \bibfield  [0]{\@secondoftwo}%
\providecommand \translation [1]{[#1]}%
\providecommand \BibitemOpen [0]{}%
\providecommand \bibitemStop [0]{}%
\providecommand \bibitemNoStop [0]{.\EOS\space}%
\providecommand \EOS [0]{\spacefactor3000\relax}%
\providecommand \BibitemShut  [1]{\csname bibitem#1\endcsname}%
\let\auto@bib@innerbib\@empty
\bibitem [{\citenamefont {Christou}\ \emph {et~al.}(2000)\citenamefont
  {Christou}, \citenamefont {Gatteschi}, \citenamefont {Hendrickson},\ and\
  \citenamefont {Sessoli}}]{christou2000single}%
  \BibitemOpen
  \bibfield  {author} {\bibinfo {author} {\bibfnamefont {G.}~\bibnamefont
  {Christou}}, \bibinfo {author} {\bibfnamefont {D.}~\bibnamefont {Gatteschi}},
  \bibinfo {author} {\bibfnamefont {D.~N.}\ \bibnamefont {Hendrickson}}, \ and\
  \bibinfo {author} {\bibfnamefont {R.}~\bibnamefont {Sessoli}},\ }\href@noop
  {} {\bibfield  {journal} {\bibinfo  {journal} {Mrs Bulletin}\ }\textbf
  {\bibinfo {volume} {25}},\ \bibinfo {pages} {66} (\bibinfo {year}
  {2000})}\BibitemShut {NoStop}%
\bibitem [{\citenamefont {Sessoli}\ \emph
  {et~al.}(1993{\natexlab{a}})\citenamefont {Sessoli}, \citenamefont {Tsai},
  \citenamefont {Schake}, \citenamefont {Wang}, \citenamefont {Vincent},
  \citenamefont {Folting}, \citenamefont {Gatteschi}, \citenamefont
  {Christou},\ and\ \citenamefont {Hendrickson}}]{sessoli1993high}%
  \BibitemOpen
  \bibfield  {author} {\bibinfo {author} {\bibfnamefont {R.}~\bibnamefont
  {Sessoli}}, \bibinfo {author} {\bibfnamefont {H.~L.}\ \bibnamefont {Tsai}},
  \bibinfo {author} {\bibfnamefont {A.~R.}\ \bibnamefont {Schake}}, \bibinfo
  {author} {\bibfnamefont {S.}~\bibnamefont {Wang}}, \bibinfo {author}
  {\bibfnamefont {J.~B.}\ \bibnamefont {Vincent}}, \bibinfo {author}
  {\bibfnamefont {K.}~\bibnamefont {Folting}}, \bibinfo {author} {\bibfnamefont
  {D.}~\bibnamefont {Gatteschi}}, \bibinfo {author} {\bibfnamefont
  {G.}~\bibnamefont {Christou}}, \ and\ \bibinfo {author} {\bibfnamefont
  {D.~N.}\ \bibnamefont {Hendrickson}},\ }\href@noop {} {\bibfield  {journal}
  {\bibinfo  {journal} {Journal of the American Chemical Society}\ }\textbf
  {\bibinfo {volume} {115}},\ \bibinfo {pages} {1804} (\bibinfo {year}
  {1993}{\natexlab{a}})}\BibitemShut {NoStop}%
\bibitem [{\citenamefont {Sessoli}\ \emph
  {et~al.}(1993{\natexlab{b}})\citenamefont {Sessoli}, \citenamefont
  {Gatteschi}, \citenamefont {Caneschi},\ and\ \citenamefont
  {Novak}}]{sessoli1993magnetic}%
  \BibitemOpen
  \bibfield  {author} {\bibinfo {author} {\bibfnamefont {R.}~\bibnamefont
  {Sessoli}}, \bibinfo {author} {\bibfnamefont {D.}~\bibnamefont {Gatteschi}},
  \bibinfo {author} {\bibfnamefont {A.}~\bibnamefont {Caneschi}}, \ and\
  \bibinfo {author} {\bibfnamefont {M.}~\bibnamefont {Novak}},\ }\href@noop {}
  {\bibfield  {journal} {\bibinfo  {journal} {Nature}\ }\textbf {\bibinfo
  {volume} {365}},\ \bibinfo {pages} {141} (\bibinfo {year}
  {1993}{\natexlab{b}})}\BibitemShut {NoStop}%
\bibitem [{\citenamefont {Tasiopoulos}\ \emph {et~al.}(2004)\citenamefont
  {Tasiopoulos}, \citenamefont {Vinslava}, \citenamefont {Wernsdorfer},
  \citenamefont {Abboud},\ and\ \citenamefont
  {Christou}}]{tasiopoulos2004giant}%
  \BibitemOpen
  \bibfield  {author} {\bibinfo {author} {\bibfnamefont {A.~J.}\ \bibnamefont
  {Tasiopoulos}}, \bibinfo {author} {\bibfnamefont {A.}~\bibnamefont
  {Vinslava}}, \bibinfo {author} {\bibfnamefont {W.}~\bibnamefont
  {Wernsdorfer}}, \bibinfo {author} {\bibfnamefont {K.~A.}\ \bibnamefont
  {Abboud}}, \ and\ \bibinfo {author} {\bibfnamefont {G.}~\bibnamefont
  {Christou}},\ }\href@noop {} {\bibfield  {journal} {\bibinfo  {journal}
  {Angewandte Chemie}\ }\textbf {\bibinfo {volume} {116}},\ \bibinfo {pages}
  {2169} (\bibinfo {year} {2004})}\BibitemShut {NoStop}%
\bibitem [{\citenamefont {Vinslava}\ \emph {et~al.}(2016)\citenamefont
  {Vinslava}, \citenamefont {Tasiopoulos}, \citenamefont {Wernsdorfer},
  \citenamefont {Abboud},\ and\ \citenamefont
  {Christou}}]{vinslava2016molecules}%
  \BibitemOpen
  \bibfield  {author} {\bibinfo {author} {\bibfnamefont {A.}~\bibnamefont
  {Vinslava}}, \bibinfo {author} {\bibfnamefont {A.~J.}\ \bibnamefont
  {Tasiopoulos}}, \bibinfo {author} {\bibfnamefont {W.}~\bibnamefont
  {Wernsdorfer}}, \bibinfo {author} {\bibfnamefont {K.~A.}\ \bibnamefont
  {Abboud}}, \ and\ \bibinfo {author} {\bibfnamefont {G.}~\bibnamefont
  {Christou}},\ }\href@noop {} {\bibfield  {journal} {\bibinfo  {journal}
  {Inorganic chemistry}\ }\textbf {\bibinfo {volume} {55}},\ \bibinfo {pages}
  {3419} (\bibinfo {year} {2016})}\BibitemShut {NoStop}%
\bibitem [{\citenamefont {Golinelli}\ \emph {et~al.}(1992)\citenamefont
  {Golinelli}, \citenamefont {Jolicoeur},\ and\ \citenamefont
  {Lacaze}}]{golinelli1992haldane}%
  \BibitemOpen
  \bibfield  {author} {\bibinfo {author} {\bibfnamefont {O.}~\bibnamefont
  {Golinelli}}, \bibinfo {author} {\bibfnamefont {T.}~\bibnamefont
  {Jolicoeur}}, \ and\ \bibinfo {author} {\bibfnamefont {R.}~\bibnamefont
  {Lacaze}},\ }\href@noop {} {\bibfield  {journal} {\bibinfo  {journal}
  {Physical Review B}\ }\textbf {\bibinfo {volume} {45}},\ \bibinfo {pages}
  {9798} (\bibinfo {year} {1992})}\BibitemShut {NoStop}%
\bibitem [{\citenamefont {Owerre}\ and\ \citenamefont
  {Paranjape}(2014)}]{owerre2014haldane}%
  \BibitemOpen
  \bibfield  {author} {\bibinfo {author} {\bibfnamefont {S.}~\bibnamefont
  {Owerre}}\ and\ \bibinfo {author} {\bibfnamefont {M.}~\bibnamefont
  {Paranjape}},\ }\href@noop {} {\bibfield  {journal} {\bibinfo  {journal}
  {Physics Letters A}\ }\textbf {\bibinfo {volume} {378}},\ \bibinfo {pages}
  {3066} (\bibinfo {year} {2014})}\BibitemShut {NoStop}%
\bibitem [{\citenamefont {Schollw{\"o}ck}\ and\ \citenamefont
  {Jolic{\oe}ur}(1995)}]{schollwock1995haldane}%
  \BibitemOpen
  \bibfield  {author} {\bibinfo {author} {\bibfnamefont {U.}~\bibnamefont
  {Schollw{\"o}ck}}\ and\ \bibinfo {author} {\bibfnamefont {T.}~\bibnamefont
  {Jolic{\oe}ur}},\ }\href@noop {} {\bibfield  {journal} {\bibinfo  {journal}
  {EPL (Europhysics Letters)}\ }\textbf {\bibinfo {volume} {30}},\ \bibinfo
  {pages} {493} (\bibinfo {year} {1995})}\BibitemShut {NoStop}%
\bibitem [{\citenamefont {Regnault}\ \emph {et~al.}(2002)\citenamefont
  {Regnault}, \citenamefont {Jolicoeur}, \citenamefont {Sessoli}, \citenamefont
  {Gatteschi},\ and\ \citenamefont {Verdaguer}}]{regnault2002exchange}%
  \BibitemOpen
  \bibfield  {author} {\bibinfo {author} {\bibfnamefont {N.}~\bibnamefont
  {Regnault}}, \bibinfo {author} {\bibfnamefont {T.}~\bibnamefont {Jolicoeur}},
  \bibinfo {author} {\bibfnamefont {R.}~\bibnamefont {Sessoli}}, \bibinfo
  {author} {\bibfnamefont {D.}~\bibnamefont {Gatteschi}}, \ and\ \bibinfo
  {author} {\bibfnamefont {M.}~\bibnamefont {Verdaguer}},\ }\href@noop {}
  {\bibfield  {journal} {\bibinfo  {journal} {Physical Review B}\ }\textbf
  {\bibinfo {volume} {66}},\ \bibinfo {pages} {054409} (\bibinfo {year}
  {2002})}\BibitemShut {NoStop}%
\bibitem [{\citenamefont {Chaboussant}\ \emph {et~al.}(2004)\citenamefont
  {Chaboussant}, \citenamefont {Sieber}, \citenamefont {Ochsenbein},
  \citenamefont {G{\"u}del}, \citenamefont {Murrie}, \citenamefont {Honecker},
  \citenamefont {Fukushima},\ and\ \citenamefont
  {Normand}}]{chaboussant2004exchange}%
  \BibitemOpen
  \bibfield  {author} {\bibinfo {author} {\bibfnamefont {G.}~\bibnamefont
  {Chaboussant}}, \bibinfo {author} {\bibfnamefont {A.}~\bibnamefont {Sieber}},
  \bibinfo {author} {\bibfnamefont {S.}~\bibnamefont {Ochsenbein}}, \bibinfo
  {author} {\bibfnamefont {H.-U.}\ \bibnamefont {G{\"u}del}}, \bibinfo {author}
  {\bibfnamefont {M.}~\bibnamefont {Murrie}}, \bibinfo {author} {\bibfnamefont
  {A.}~\bibnamefont {Honecker}}, \bibinfo {author} {\bibfnamefont
  {N.}~\bibnamefont {Fukushima}}, \ and\ \bibinfo {author} {\bibfnamefont
  {B.}~\bibnamefont {Normand}},\ }\href@noop {} {\bibfield  {journal} {\bibinfo
   {journal} {Physical Review B}\ }\textbf {\bibinfo {volume} {70}},\ \bibinfo
  {pages} {104422} (\bibinfo {year} {2004})}\BibitemShut {NoStop}%
\bibitem [{\citenamefont {Bagai}\ and\ \citenamefont
  {Christou}(2009)}]{bagai2009drosophila}%
  \BibitemOpen
  \bibfield  {author} {\bibinfo {author} {\bibfnamefont {R.}~\bibnamefont
  {Bagai}}\ and\ \bibinfo {author} {\bibfnamefont {G.}~\bibnamefont
  {Christou}},\ }\href@noop {} {\bibfield  {journal} {\bibinfo  {journal}
  {Chemical Society Reviews}\ }\textbf {\bibinfo {volume} {38}},\ \bibinfo
  {pages} {1011} (\bibinfo {year} {2009})}\BibitemShut {NoStop}%
\bibitem [{\citenamefont {Park}\ \emph {et~al.}(2004)\citenamefont {Park},
  \citenamefont {Pederson},\ and\ \citenamefont
  {Hellberg}}]{park2004properties}%
  \BibitemOpen
  \bibfield  {author} {\bibinfo {author} {\bibfnamefont {K.}~\bibnamefont
  {Park}}, \bibinfo {author} {\bibfnamefont {M.~R.}\ \bibnamefont {Pederson}},
  \ and\ \bibinfo {author} {\bibfnamefont {C.~S.}\ \bibnamefont {Hellberg}},\
  }\href@noop {} {\bibfield  {journal} {\bibinfo  {journal} {Physical Review
  B}\ }\textbf {\bibinfo {volume} {69}},\ \bibinfo {pages} {014416} (\bibinfo
  {year} {2004})}\BibitemShut {NoStop}%
\bibitem [{\citenamefont {M{\"u}ller}\ \emph {et~al.}(2001)\citenamefont
  {M{\"u}ller}, \citenamefont {Luban}, \citenamefont {Schr{\"o}der},
  \citenamefont {Modler}, \citenamefont {K{\"o}gerler}, \citenamefont
  {Axenovich}, \citenamefont {Schnack}, \citenamefont {Canfield}, \citenamefont
  {Bud'ko},\ and\ \citenamefont {Harrison}}]{muller2001classical}%
  \BibitemOpen
  \bibfield  {author} {\bibinfo {author} {\bibfnamefont {A.}~\bibnamefont
  {M{\"u}ller}}, \bibinfo {author} {\bibfnamefont {M.}~\bibnamefont {Luban}},
  \bibinfo {author} {\bibfnamefont {C.}~\bibnamefont {Schr{\"o}der}}, \bibinfo
  {author} {\bibfnamefont {R.}~\bibnamefont {Modler}}, \bibinfo {author}
  {\bibfnamefont {P.}~\bibnamefont {K{\"o}gerler}}, \bibinfo {author}
  {\bibfnamefont {M.}~\bibnamefont {Axenovich}}, \bibinfo {author}
  {\bibfnamefont {J.}~\bibnamefont {Schnack}}, \bibinfo {author} {\bibfnamefont
  {P.}~\bibnamefont {Canfield}}, \bibinfo {author} {\bibfnamefont
  {S.}~\bibnamefont {Bud'ko}}, \ and\ \bibinfo {author} {\bibfnamefont
  {N.}~\bibnamefont {Harrison}},\ }\href@noop {} {\bibfield  {journal}
  {\bibinfo  {journal} {ChemPhysChem}\ }\textbf {\bibinfo {volume} {2}},\
  \bibinfo {pages} {517} (\bibinfo {year} {2001})}\BibitemShut {NoStop}%
\bibitem [{\citenamefont {Exler}\ and\ \citenamefont
  {Schnack}(2003)}]{exler2003evaluation}%
  \BibitemOpen
  \bibfield  {author} {\bibinfo {author} {\bibfnamefont {M.}~\bibnamefont
  {Exler}}\ and\ \bibinfo {author} {\bibfnamefont {J.}~\bibnamefont
  {Schnack}},\ }\href@noop {} {\bibfield  {journal} {\bibinfo  {journal}
  {Physical Review B}\ }\textbf {\bibinfo {volume} {67}},\ \bibinfo {pages}
  {094440} (\bibinfo {year} {2003})}\BibitemShut {NoStop}%
\bibitem [{\citenamefont {Neuscamman}\ and\ \citenamefont
  {Chan}(2012)}]{neuscamman2012correlator}%
  \BibitemOpen
  \bibfield  {author} {\bibinfo {author} {\bibfnamefont {E.}~\bibnamefont
  {Neuscamman}}\ and\ \bibinfo {author} {\bibfnamefont {G.~K.-L.}\ \bibnamefont
  {Chan}},\ }\href@noop {} {\bibfield  {journal} {\bibinfo  {journal} {Physical
  Review B}\ }\textbf {\bibinfo {volume} {86}},\ \bibinfo {pages} {064402}
  (\bibinfo {year} {2012})}\BibitemShut {NoStop}%
\bibitem [{\citenamefont {Ummethum}\ \emph {et~al.}(2013)\citenamefont
  {Ummethum}, \citenamefont {Schnack},\ and\ \citenamefont
  {L{\"a}uchli}}]{ummethum2013large}%
  \BibitemOpen
  \bibfield  {author} {\bibinfo {author} {\bibfnamefont {J.}~\bibnamefont
  {Ummethum}}, \bibinfo {author} {\bibfnamefont {J.}~\bibnamefont {Schnack}}, \
  and\ \bibinfo {author} {\bibfnamefont {A.~M.}\ \bibnamefont {L{\"a}uchli}},\
  }\href@noop {} {\bibfield  {journal} {\bibinfo  {journal} {Journal of
  Magnetism and Magnetic Materials}\ }\textbf {\bibinfo {volume} {327}},\
  \bibinfo {pages} {103} (\bibinfo {year} {2013})}\BibitemShut {NoStop}%
\bibitem [{\citenamefont {Kohn}\ and\ \citenamefont
  {Sham}(1965)}]{kohn1965self}%
  \BibitemOpen
  \bibfield  {author} {\bibinfo {author} {\bibfnamefont {W.}~\bibnamefont
  {Kohn}}\ and\ \bibinfo {author} {\bibfnamefont {L.~J.}\ \bibnamefont
  {Sham}},\ }\href@noop {} {\bibfield  {journal} {\bibinfo  {journal} {Physical
  review}\ }\textbf {\bibinfo {volume} {140}},\ \bibinfo {pages} {A1133}
  (\bibinfo {year} {1965})}\BibitemShut {NoStop}%
\bibitem [{\citenamefont {Perdew}\ \emph {et~al.}(1996)\citenamefont {Perdew},
  \citenamefont {Burke},\ and\ \citenamefont
  {Ernzerhof}}]{perdew1996generalized}%
  \BibitemOpen
  \bibfield  {author} {\bibinfo {author} {\bibfnamefont {J.~P.}\ \bibnamefont
  {Perdew}}, \bibinfo {author} {\bibfnamefont {K.}~\bibnamefont {Burke}}, \
  and\ \bibinfo {author} {\bibfnamefont {M.}~\bibnamefont {Ernzerhof}},\
  }\href@noop {} {\bibfield  {journal} {\bibinfo  {journal} {Physical review
  letters}\ }\textbf {\bibinfo {volume} {77}},\ \bibinfo {pages} {3865}
  (\bibinfo {year} {1996})}\BibitemShut {NoStop}%
\bibitem [{\citenamefont {Bl{\"o}chl}(1994)}]{blochl1994projector}%
  \BibitemOpen
  \bibfield  {author} {\bibinfo {author} {\bibfnamefont {P.~E.}\ \bibnamefont
  {Bl{\"o}chl}},\ }\href@noop {} {\bibfield  {journal} {\bibinfo  {journal}
  {Physical review B}\ }\textbf {\bibinfo {volume} {50}},\ \bibinfo {pages}
  {17953} (\bibinfo {year} {1994})}\BibitemShut {NoStop}%
\bibitem [{\citenamefont {Kresse}\ and\ \citenamefont
  {Joubert}(1999)}]{kresse1999ultrasoft}%
  \BibitemOpen
  \bibfield  {author} {\bibinfo {author} {\bibfnamefont {G.}~\bibnamefont
  {Kresse}}\ and\ \bibinfo {author} {\bibfnamefont {D.}~\bibnamefont
  {Joubert}},\ }\href@noop {} {\bibfield  {journal} {\bibinfo  {journal}
  {Physical Review B}\ }\textbf {\bibinfo {volume} {59}},\ \bibinfo {pages}
  {1758} (\bibinfo {year} {1999})}\BibitemShut {NoStop}%
\bibitem [{\citenamefont {Kresse}\ and\ \citenamefont
  {Furthm{\"u}ller}(1996{\natexlab{a}})}]{kresse1996efficiency}%
  \BibitemOpen
  \bibfield  {author} {\bibinfo {author} {\bibfnamefont {G.}~\bibnamefont
  {Kresse}}\ and\ \bibinfo {author} {\bibfnamefont {J.}~\bibnamefont
  {Furthm{\"u}ller}},\ }\href@noop {} {\bibfield  {journal} {\bibinfo
  {journal} {Computational materials science}\ }\textbf {\bibinfo {volume}
  {6}},\ \bibinfo {pages} {15} (\bibinfo {year}
  {1996}{\natexlab{a}})}\BibitemShut {NoStop}%
\bibitem [{\citenamefont {Kresse}\ and\ \citenamefont
  {Furthm{\"u}ller}(1996{\natexlab{b}})}]{kresse1996efficient}%
  \BibitemOpen
  \bibfield  {author} {\bibinfo {author} {\bibfnamefont {G.}~\bibnamefont
  {Kresse}}\ and\ \bibinfo {author} {\bibfnamefont {J.}~\bibnamefont
  {Furthm{\"u}ller}},\ }\href@noop {} {\bibfield  {journal} {\bibinfo
  {journal} {Physical review B}\ }\textbf {\bibinfo {volume} {54}},\ \bibinfo
  {pages} {11169} (\bibinfo {year} {1996}{\natexlab{b}})}\BibitemShut {NoStop}%
\bibitem [{\citenamefont {Schnack}\ and\ \citenamefont
  {Luban}(2000)}]{schnack}%
  \BibitemOpen
  \bibfield  {author} {\bibinfo {author} {\bibfnamefont {J.}~\bibnamefont
  {Schnack}}\ and\ \bibinfo {author} {\bibfnamefont {M.}~\bibnamefont
  {Luban}},\ }\href@noop {} {\bibfield  {journal} {\bibinfo  {journal}
  {Physical Review B}\ }\textbf {\bibinfo {volume} {63}},\ \bibinfo {pages}
  {014418} (\bibinfo {year} {2000})}\BibitemShut {NoStop}%
\end{thebibliography}%

\end{document}